# Mobility anisotropy in monolayer black phosphorus due to charged impurities

Yue Liu[1], Tony Low[1,*] and P. Paul Ruden[1]

[1]*Department of Electrical and Computer Engineering, University of Minnesota, Minneapolis, Minnesota, 55455, USA*

ABSTRACT

We study the charged impurity limited mobility in black phosphorus, a highly anisotropic layered material. We compute the mobility within the Boltzmann transport equation under detailed balance condition, and taking into account the anisotropy in transport and electronic structure. For carrier densities accessible in experiments, we obtained an anisotropy ratio of 3 ~ 4 at zero temperature, two-folds larger than that observed in experiments on multilayers samples. We discuss also how the anisotropy depends on carrier density and impurity distribution.

*Author to whom correspondence should be addressed: tlow@umn.edu

## I. Introduction

Black phosphorus (BP), its puckered layered form, is one of the thermodynamically more stable phases of phosphorus at ambient temperature and pressure.[1-4] The successful exfoliation of BP[5-6] multilayers and monolayers has triggered tremendous interest in this material. In particular, multilayer BP has a direct bandgap that spans from $0.3 \sim 1.5 eV$ depending on the number of layers,[7-10] hence making it an excellent candidate for infrared optoelectronics[11-13]. Its good electrical transport properties, with highest carrier mobility after graphene and a finite electronic gap larger than $k_B T$ at ambient, make it a promising candidate for nanoelectronics.[15,16] Since each BP layer has a puckered structure due to $sp^3$ hybridization, it also reveals highly anisotropic electrical and optical properties. This includes anisotropic charge carrier mobility[4,5,14,17] and linear dichroism in optical absorption spectra.[2,8-10,14,18] Indeed, the BP crystal structure leads to highly anisotropic energy bands where the in-plane effective masses along the two crystal axes, armchair and zigzag, can differ by an order of magnitude. For example, the effective masses in bulk BP were measured to be $m_{xx}^e = m_{xx}^h = 0.08 m_0$ and $m_{yy}^e = 0.7 m_0$, $m_{yy}^h = 1.0 m_0$ with cyclotron resonance techniques.[19] In monolayer BP, these masses were predicted to be $m_{xx}^e = m_{xx}^h = 0.15 m_0$, with $m_{yy}^e$ and $m_{yy}^h$ being same as in bulk BP.[8,20] However, low temperature measurement on multilayer BP thus far only have yielded anisotropy ratios of between 1.5 to 2[6,14,17]. Reconciling this ratio to the anisotropy in its electronic bands is important in the understanding of its relevant scattering mechanisms and electronic transport properties.

Although there have been several theoretical studies on the anisotropic transport properties of BP[9,21,22], the anisotropy of scattering was not treated explicitly. This is because the momentum relaxation time of an anisotropic medium, solved with the Boltzmann transport equation, does not have a closed form solution.[23,24] That problem has been avoided so far with approximations that are not entirely suitable for anisotropic materials. In this paper, we describe our methodology of solving for the charged-impurity-limited anisotropic mobility of BP with the Boltzmann transport equation within the relaxation time approximation, accounting for the full anisotropy of the problem. The momentum relaxation time depends not only on the direction of electric field, but also the wave-vector of incoming state. We found also that the anisotropic mobility ratio is highly sensitive to the distribution of the charged impurities, and can range from 1.5 to 7. For uniformly distributed impurities in the substrate, we estimated an anisotropic mobility ratio of about $\sim 3.5 - 4$ for monolayer BP. We discuss how mobility and its anisotropy ratio depend on impurity distance and carrier density.

## II. Model description

We consider monolayer BP on an insulating substrate as illustrated in Fig. 1a, with charged impurities distributed at some distance $d$ from the interface interacting with charge carriers in the BP via Coulomb interaction. We model the anisotropic energy dispersion of BP in the vicinity of the conduction band minimum at the $\Gamma$ valley as follows,[1,2]

$$E(\vec{k}) = \frac{\hbar^2}{2}\left(\frac{1}{m_{xx}}k_x^2 + \frac{1}{m_{yy}}k_y^2\right) \quad (1)$$

, where $m_{xx}$ and $m_{yy}$ are effective masses along $x$ (armchair) and $y$ (zigzag) direction. Coulomb scattering is an elastic process, and Fig. 1b illustrates the scattering phase space from the incoming $|\vec{k}_i\rangle$ to the outgoing $|\vec{k}_j\rangle$ states. The transition rate $P_{\vec{k}_i,\vec{k}_j}$ can be expressed by the Fermi's golden rule,

$$P_{\vec{k}_i,\vec{k}_j} = \frac{2\pi}{\hbar}|\langle\vec{k}_j|H|\vec{k}_i\rangle|^2 n_{imp}\delta[E(\vec{k}_i) - E(\vec{k}_j)] \quad (2)$$

$H$ is the interaction Hamiltonian describing the Coulomb perturbation. In this work, we assume a typical impurity concentration of $n_{imp} = 10^{12} cm^{-2}$ unless stated otherwise. The constant density-of-states (DOS) of BP, $g_{2D} = \frac{m_d}{\pi\hbar^2}$, where the DOS effective mass $m_d = \sqrt{m_{xx}m_{yy}}$, allows for a linear, static Thomas-Fermi screening description, as it is commonly employed for a conventional 2D electron gas,[25]

$$|\langle\vec{k}_j|H|\vec{k}_i\rangle| = \frac{2\pi e^2 e^{-q\cdot d}}{q\kappa + 2\pi e^2 \Pi(\vec{q})} \quad (3)$$

The effective dielectric constant is $\kappa \approx 2.5$ for the air and SiO$_2$ (substrate) half spaces. The scattering wave vector is denoted by $\vec{q} = \vec{k}_j - \vec{k}_i$ and $\Pi(\vec{q})$ is the anisotropic 2D polarizability of BP. In general, $\Pi(\vec{q})$ depends also on the direction of $\vec{q}$, and in the limit of zero temperature reduces to,[26]

$$\Pi(\vec{q}) = g_{2D} Re\left[1 - \sqrt{1 - \frac{4E_F}{E(\vec{q})}}\right] \quad (4)$$

, where $E_F = \frac{\hbar^2 \pi n}{m_d}$ is Fermi energy, and $n$ is the carrier density in BP. Furthermore, it can be shown that $\Pi(\vec{q})$ is isotropic for $q \equiv |\vec{q}|$ constrained by $q \leq 2|\vec{k}_F \cdot \hat{q}|$, which is also the phase space coinciding with the scattering illustrated in Fig. 1b. Here $\Pi(\vec{q})$ simply reduces to $g_{2D}$ in the zero temperature limit.[26]

We are now ready to write down the anisotropic momentum relaxation time, $\tau_m$, which is the time needed for the momentum distribution to relaxed, derived from the detailed balance Boltzmann transport equation (See Supplementary Information.).[27]

$$\frac{1}{\tau_m(\hat{\xi},\vec{k}_i)} = \frac{1}{(2\pi)^2}\int_{all\ \vec{k}_j} d^2\vec{k}_j\ P_{\vec{k}_i,\vec{k}_j}\left\{1 - \frac{[\hat{\xi}\cdot\vec{v}(\vec{k}_j)]\tau_m(\hat{\xi},\vec{k}_j)}{[\hat{\xi}\cdot\vec{v}(\vec{k}_i)]\tau_m(\hat{\xi},\vec{k}_i)}\right\} \tag{5}$$

$\hat{\xi}$ is the direction of the applied electric field and $\vec{v}(\vec{k}) = \frac{1}{\hbar}\nabla_{\vec{k}}E(\vec{k})$ is the group velocity. For $T = 0K$, scattering only occurs at the Fermi level, that is $E(\vec{k}_i) = E_F$ in Eq. 2. It is worth pointing out that the relaxation time of an anisotropic material depends both on the direction of the electric field, $\hat{\xi}$, and on the wave-vector of incoming state $|\vec{k}_i\rangle$. The magnitude of the electric field $|\vec{\xi}|$ is irrelevant in linear response.

For an isotropic 2D electron gas material, such a GaAs heterostructure, $m_{xx} = m_{yy}$, and the relaxation time is the same for all $|\vec{k}_i\rangle$ and does not depend on $\hat{\xi}$, i.e. $\tau_m(\hat{\xi},\vec{k}_i) = \tau_{iso}$. Then Eq. 5 can be reduced to an explicit integral:

$$\frac{1}{\tau_{iso}} = \frac{1}{(2\pi)^2}\int_{all\ \vec{k}_j} d^2\vec{k}_j\ P_{\vec{k}_i,\vec{k}_j}(1 - cos\theta_{ij}) \tag{6}$$

, where $\theta_{ij}$ is the scattering angle between $|\vec{k}_i\rangle$ and $|\vec{k}_j\rangle$. Eq. 6 can be used as a sanity check for the calculation (See Supplementary Information).[27]

The strongly anisotropic electronic structure of BP $\left(\frac{m_{yy}}{m_{xx}} \approx 7\right)$ requires solving of the momentum relaxation time $\tau_m(\hat{\xi},\vec{k}_i)$ from the implicit integral in Eq. 5. We discuss briefly our numerical procedure. First, we recast Eq. 5 in cylindrical coordinates, arriving at,

$$\frac{1}{\tau_m(\hat{\xi},\vec{k}_i)} = \frac{1}{(2\pi)^2}\int_0^{2\pi} d\theta_j \left\{\beta M_{i,j}\frac{1}{2\beta}\left[1 - \frac{\alpha[\vec{k}_j = (\beta,\theta_j)]\tau_m[\hat{\xi},\vec{k}_j = (\beta,\theta_j)]}{\alpha(\vec{k}_i)\tau_m(\hat{\xi},\vec{k}_i)}\right]\right\} \tag{7}$$

, where

$$M_{i,j} = \frac{2\pi}{\hbar}|\langle\vec{k}_j|H|\vec{k}_i\rangle|^2 n_{imp}\left(\frac{\hbar^2 cos^2\theta_j}{2m_{xx}} + \frac{\hbar^2 sin^2\theta_j}{2m_{yy}}\right)^{-1} \tag{8}$$

, and we introduce $\beta = k_j \sqrt{\frac{E(\vec{k}_i)}{E(\vec{k}_j)}}$, $\alpha(\vec{k}) = \hat{\xi} \cdot \vec{v}(\vec{k})$ to simplify notations. We can further identify $W(\theta_j) = \left(\frac{\hbar^2 \cos^2\theta_j}{2m_{xx}} + \frac{\hbar^2 \sin^2\theta_j}{2m_{yy}}\right)^{-1}$, which can be viewed as a DOS along the energy contour. In arriving at Eq. 7, we made use of the delta function property, $\delta(\beta^2 - k_j^2) = \frac{1}{2\beta_j}[\delta(k_j + \beta) + \delta(k_j - \beta)]$. Second, we rewrite Eq. 7 as a system of linear equations of $\tau_i \equiv \tau_m(\hat{\xi}, \vec{k}_i)$, where the index $i = 1, \ldots, N$ stands for the $i$th discrete point of the incoming state $|\vec{k}_i\rangle$ along the elliptical $k$-contour. Writing $b = \frac{1}{(2\pi)^2}\frac{\Delta\theta_j}{2}$, with $\Delta\theta_j = \frac{2\pi}{N-1}$, and $\alpha_i \equiv \alpha(\vec{k}_i)$, Eq. 7 can then be reformulated as,

$$-\alpha_i + b \sum_j M_{i,j}\,\alpha_i \tau_i = b(M_{i,1}\alpha_1 \quad M_{i,2}\alpha_2 \quad \cdots \quad M_{i,N}\alpha_N)\begin{pmatrix}\tau_1 \\ \tau_2 \\ \vdots \\ \tau_N\end{pmatrix} \quad (9)$$

for all $i$. Eq. 9 can be written in matrix form as $[T]|\tau_m\rangle = |\alpha\rangle$, the array of $\tau_m$ can be solved by inverting $[T]$ iteratively. In our computation, we employed a total of $N = 1000$ points for our discretization.

### III. Anisotropic momentum relaxation time

Following the model described in section II, we compute the anisotropic momentum relaxation times $\tau_m(\hat{x}, \vec{k}_i)$ and $\tau_m(\hat{y}, \vec{k}_i)$, assuming $T = 0K$, and hole effective masses of $m_{xx} = 0.15 m_0$ and $m_{yy} = 1.0 m_0$ unless otherwise stated.[8] The results are plotted in Fig. 2a. The averaged momentum relaxation times, $\langle\tau_m\rangle = \frac{1}{N}\sum_i \tau_m(\hat{\xi}, \vec{k}_i)$, are plotted in Fig. 2b. The calculated momentum relaxation time is on the order of picoseconds for the assumed impurity concentration with $\tau_m(\hat{y}, \vec{k}_i) > \tau_m(\hat{x}, \vec{k}_i)$ and the relaxation time anisotropy ratio of $\approx 5$ is obtained. Momentum relaxation favors back-scattering against the direction of the electric field. This is apparent from the $1 - \frac{[\hat{\xi}\cdot\vec{v}(\vec{k}_j)]\tau_m(\hat{\xi},\vec{k}_j)}{[\hat{\xi}\cdot\vec{v}(\vec{k}_i)]\tau_m(\hat{\xi},\vec{k}_i)}$ term in Eq. 5, which in the isotropic case will reduce to $1 - \cos\theta_{ij}$ in Eq. 6. Due to the band anisotropy, back-scattering acquires a larger $q$ when $\hat{\xi}$ is along $y$. Since $M_{i,j}$ decreases with increasing $q$, it leads to a larger momentum relaxation time, i.e. $\tau_m(\hat{y}, \vec{k}_i) > \tau_m(\hat{x}, \vec{k}_i)$.

In addition to the direction of electric field, $\tau_m$ also depends on the incoming state wavevector, $\vec{k}_i$. We consider the simpler case of $d = 0$. In this limit, the scattering matrix element $M_{i,j}$ will be independent of the initial state $\vec{k}_i$ if $q \ll \frac{2\pi e^2}{\kappa} g_{2D}$. Since $q$ increases with the Fermi energy, one can identify a carrier density at which $\tau_m$ cross-over from being independent of $\vec{k}_i$ to $\vec{k}_i$-dependent. Fig. 2a reflects this behavior. When the carrier density is small enough, such that $q$ is negligible compared to the screening term, $\frac{2\pi e^2}{\kappa} g_{2D}$, $\tau_m(\hat{\xi}, \vec{k}_i)$ is independent of $\vec{k}_i$ (red curves in Fig. 2a). Whereas for large $n$, we observe that $\tau_m$ has minima when $\theta_i = \frac{\pi}{2}$ and $\frac{3\pi}{2}$. The scattering matrix element $M_{i,j}$ in Eq. 8 depends on an effective angular DOS, $W(\theta_j)$. In the limit of extreme anisotropy, i.e. $m_{yy} \gg m_{xx}$, the maxima of $W(\theta_j)$ occur near $\theta_j = \frac{\pi}{2}$ and $\frac{3\pi}{2}$. On the other hand, in the isotropic limit with $m_{yy} = m_{xx}$, $W(\theta_j)$ is a constant. It can be seen from Fig. 1b that $q$ is zero when $\theta_i = \frac{\pi}{2}$, and the $M_{i,j}$ reaches maximum. As a result, $\tau_m$ has a minimum at $\theta_i = \frac{\pi}{2}$ for high carrier densities. Increasing carrier density $n$ increases the effective $q$ involved in scattering. Therefore increasing $d$ and/or $n$ lead to smaller $M_{i,j}$ and larger $\langle \tau_m \rangle$, as shown in Fig. 2b.

## IV. Anisotropic mobility

With $\tau_m(\hat{\xi}, \vec{k}_i)$ computed, the effective mobility $\mu$ or conductivity $\sigma$ at $T = 0K$ can be investigated. For $\hat{\xi}$ along the $x$ direction we define,

$$\mu_{xx} = \frac{g_s e}{(2\pi)^2 n} \int d^2\vec{k}\, v_x^2(\vec{k}) \tau_m(\hat{x}, \vec{k}) \frac{\partial f}{\partial E} \qquad (10)$$

$\mu_{xx}$ and $\mu_{yy}$ are non-zero elements of the mobility tensor, $v_x(\vec{k}) = \frac{\hbar k_x}{m_{xx}}$, $g_s = 2$ is the spin degeneracy, $\frac{\partial f}{\partial E} = \frac{1}{k_B T} f(E)[1 - f(E)]$ is a delta function at zero temperature, where $f(E)$ is the Fermi-Dirac distribution. As $d$ or $n$ increase, $\tau_m$ increases as previously discussed, leading to increasing mobility as shown in Fig. 3a. Although $\tau_m(\hat{x}, \vec{k}_i) < \tau_m(\hat{y}, \vec{k}_i)$, we find $\mu_{xx} > \mu_{yy}$. This is because the mobility depends on $(\hat{\xi} \cdot \vec{v})^2$, the mobility anisotropy ratio is $\frac{\mu_{xx}}{\mu_{yy}} \sim \left(\frac{m_{yy}}{m_{xx}}\right)^2 \frac{\langle \tau_m(\hat{x},\vec{k}_i)\rangle}{\langle \tau_m(\hat{y},\vec{k}_i)\rangle}$. In this case, $\left(\frac{m_{yy}}{m_{xx}}\right)^2$ is approximately 44. The smaller anisotropy ratio observed is due to the opposing trend of the momentum relaxation time. As illustrated by dashed lines in Fig. 3b, only when $d = 0$ does the anisotropy ratio decrease monotonically as a function of $n$ in the range investigated. When $d \neq 0$, for each dashed $\frac{\mu_{xx}}{\mu_{yy}}$ curve, there is a minimum located around $n_{cut} = (2\pi d^2)^{-1}$, which originates from the scattering matrix element $M_{i,j}$ depending exponentially on $q \cdot d$. Therefore $d$ defines an effective cutoff for $q$, and correspondingly a finite $n_{cut}$. Besides, all curves approximately share a similar minimum of $\approx 1.5$. For larger $d$, the minimum at $\sim(2\pi d^2)^{-1}$ is found at smaller $n$. It is interesting to note that our calculated anisotropic ratio can varies from 1.5 to 7 depending on the values of $n$ and $d$.

To evaluate an average effect of $d$, we employed a uniform impurity distribution model, which is probably the situation in experiment. Using $n_0 = \frac{n_{imp}}{t}$, to replace $n_{imp}$ in Eq. 2, $t = 300nm$ is the total thickness of the SiO$_2$ substrate, as shown in Fig. 1a, we replace the matrix element in Eq. 3 with $|\langle \vec{k}_j|H|\vec{k}_i\rangle| = \frac{2\pi e^2 e^{-q \cdot z}}{q\kappa + 2\pi e^2 \Pi(\vec{q})}$. The mobility and anisotropy ratio calculated using the uniform impurity distribution model is shown as black curves in Fig. 3a and b. As we stated above, increasing distance leads to increasing mobility. The curve varies most similarly to the $d = 0$ case and is now also monotonic as a function of $n$, which can be understood as that impurities nearest to BP have the largest influence. Uniform redistribution of impurities also reduces the sensitivity with $n$.

## V. Comparison with Experiments

In our theoretical calculation for monolayer BP, we find hole mobilities on the order of $10^3 - 10^4 cm^2/Vs$ for carrier densities $10^{12} - 10^{13} cm^{-2}$ for transport along the low effective mass direction. On the experimental side, hole Hall mobility of order $10^3 cm^2/Vs$ at carrier density of about $6.7 \times 10^{12} cm^{-2}$ has been observed at low temperature.[14,17] This would suggests that our assumed impurity concentration of $n_{imp} = 10^{12} cm^{-2}$ probably under-estimates the experimental situation or there could be other sources of scattering, e.g. neutral impurities, short-range traps states and surface roughness. A better quantity for comparison with experiment may be the anisotropy ratio, $\frac{\mu_{xx}}{\mu_{yy}}$, which does not depends on $n_{imp}$.

The mobility anisotropy ratios for holes is evaluated to be~3.5-4, across the range of hole densities shown in Fig. 3b. Using $m_{xx}^e = 0.15 m_0$ and $m_{yy}^e = 0.7 m_0$, we calculate the electron mobility anisotropy ratio as ~2.4-3.2, across the same range of carrier densities. These anisotropy ratios are larger than $\frac{\mu_{xx}}{\mu_{yy}} \approx 1.8$ obtained from Hall hole mobility measurements at $120K$[14], and that obtained from nonlocal resistance measurements, which yielded $\frac{\mu_{xx}}{\mu_{yy}} \approx 1.66 \pm 1.1$ at $5-50K$. On the other hand, angle-resolved field effect mobility measurement yield a ratio of ~2-4[17] instead.

One may identify two possible reasons for the discrepancy. First, few-layer BP samples with a thickness around $10nm$ were used in experiments. Their multi-subbands electronic structures can lead to scattering between subbands. Experimental measurements were conducted at $10-120K$, where electron-phonon scattering is not completely quenched, and might reduce the anisotropy. Moreover, the mobility anisotropy observed in experiments so far are relatively insensitive to carrier concentration, more consistent with our considered case of uniform impurity distribution. This might suggest that Coulomb impurities in experimental samples are probably due to bulk dopants, probably introduced as grown. With advancement in BP growth techniques, one may eventually approach situations where the mobility is limited by interfacial impurities like in the case of state-of-the-art semiconductor devices. We defer the study of these issues to future work. We can further try a similar calculation with short range potential, i.e. by replacing Eq. 3 with a constant scatterers, and a mobility anisotropy ratio of ~4.67 can be found, which is independent of $n$. Our study points to other source of scattering that suppress the measured anisotropy ratio in experiments.

To sum up, we have calculated the charged-impurity-scattering limited hole mobility of monolayer BP within a detailed balance Boltzmann transport model, considering the full anisotropy of the transport and electronic structure explicitly. We elucidate on the momentum relaxation time dependence on the electric field direction as well as on the incoming wave-vector. Although $\tau_m(\hat{x}, \vec{k}_i) < \tau_m(\hat{y}, \vec{k}_i)$, we find $\mu_{xx} > \mu_{yy}$ with anisotropy mobility ratios ~3.5-4. Influence of effective mass difference is compensated by the opposite trend of relaxation time. The approach outlined in this paper can also be applied to other emerging anisotropic 2D materials, such as the 1T phase of transition metal dichalcogenides[29-31] and transition metal trichalcogenides[32-34].

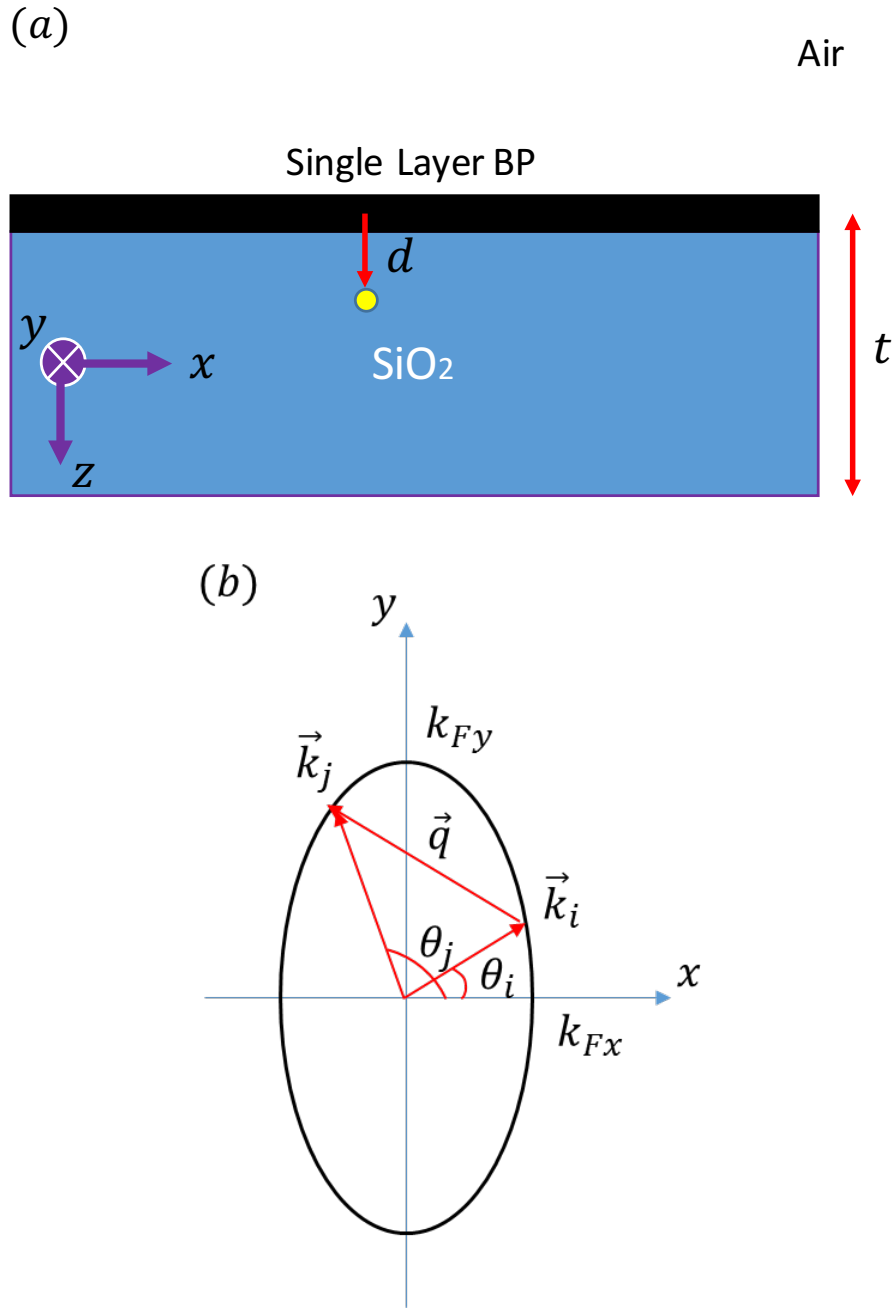

**Fig. 1.** (color online) Model structure schematic. (a) BP/SiO$_2$ as semiconductor/insulator layer structure, with an charged impurity represented by a yellow dot with distance $d$. (b) Charged impurity scattering occurs on an ellipsoidal phase space contour, with $x$ and $y$ being the armchair and zigzag directions of BP.

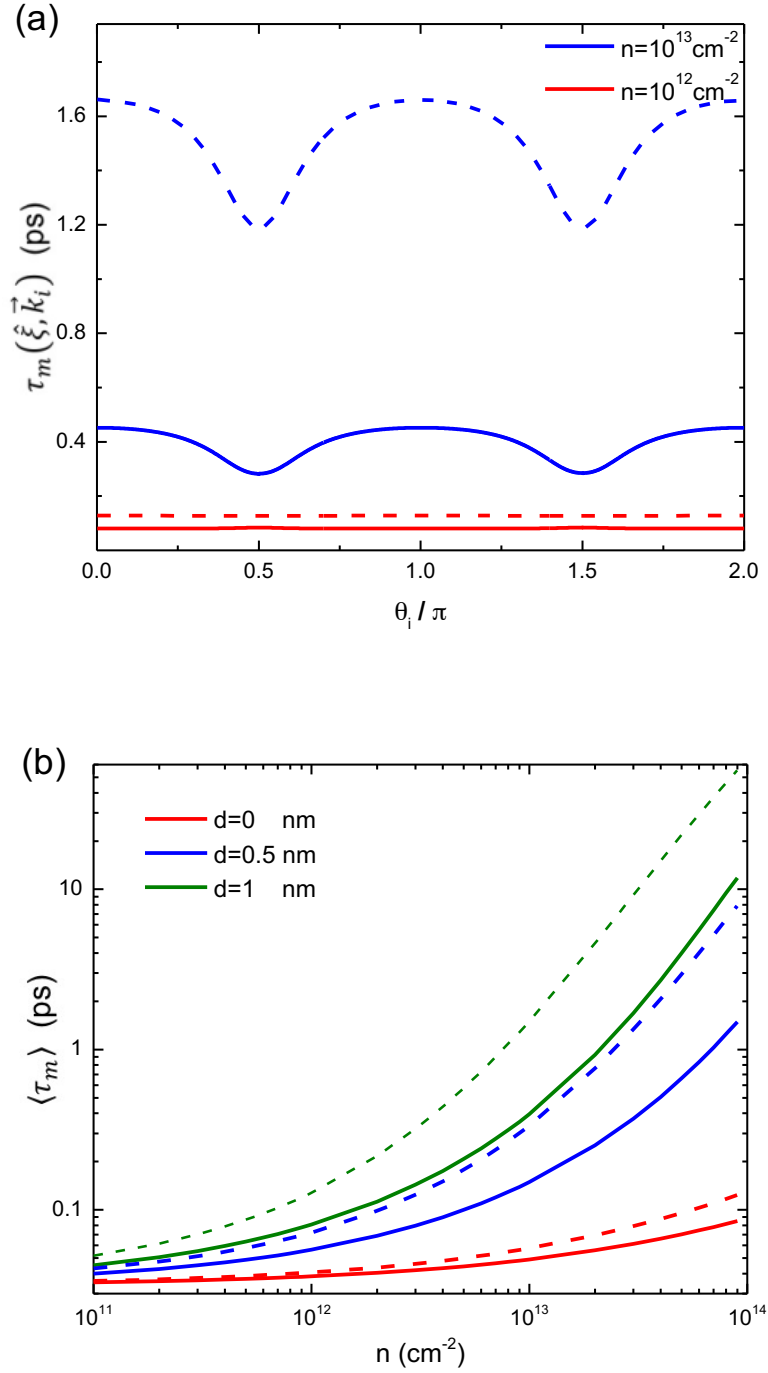

**Fig. 2.** (color online) Calculated momentum relaxation time $\tau_m$ and dependence on impurity distance, $d$, carrier density, $n$, and electric field direction, $\hat{\xi}$. Solid and dashed lines represent $\hat{\xi}$ along $x$ and $y$ directions, respectively. (a) $\tau_m$ varies with the initial wavevector, $\vec{k}_i$, for $d=1\,nm$, $\tau_m(\hat{x},\vec{k}_i) < \tau_m(\hat{y},\vec{k}_i)$. (b) Average $\langle\tau_m\rangle$ dependence on $n$, for different $d$.

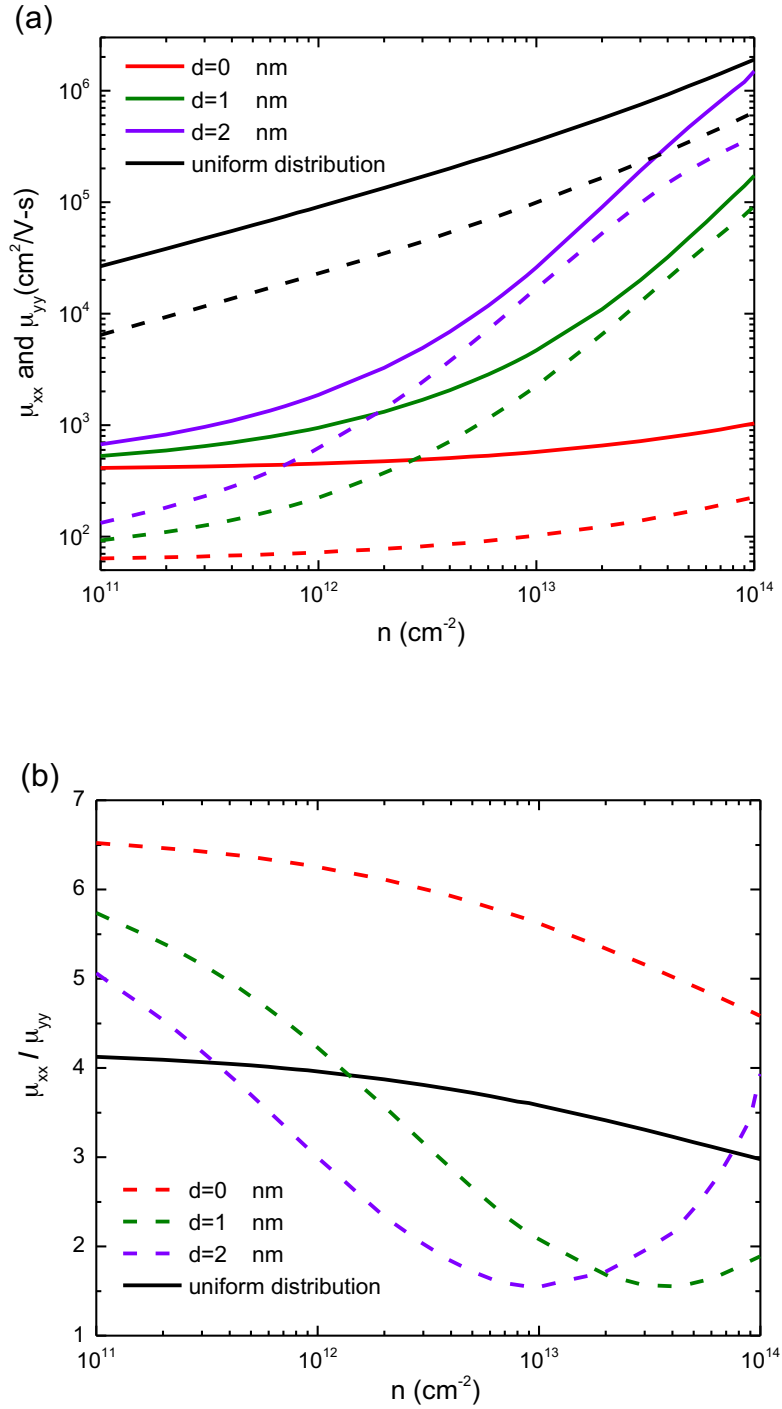

**Fig. 3**. (color online) Calculated mobility and anisotropy ratio, related to impurity distance, $d$, and carrier density, $n$. Impurity density, $n_{imp} = 10^{12} cm^{-2}$ (a) mobility $\mu_{xx} < \mu_{yy}$. Solid and dashed lines represent $\mu_{xx}$ and $\mu_{yy}$, respectively. (b) anisotropy ratio, $\frac{\mu_{xx}}{\mu_{yy}}$, changing as a function of $n$. Solid and dashed lines represent uniform model and constant distance model, respectively